\documentclass[9pt,twocolumn,twoside]{opticajnl}
\journal{opticajournal}

\setboolean{shortarticle}{false}

\usepackage{amsmath,amssymb}
\usepackage{bm}

\title{Sample Motion for Structured Illumination Fluorescence Microscopy}

\author[1,4,*]{Ruiming Cao}
\author[2,3]{Guanghan Meng}
\author[2]{Laura Waller}

\affil[1]{Department of Bioengineering, University of California, Berkeley, Berkeley CA, 94720}
\affil[2]{Department of Electrical Engineering and Computer Sciences, University of California, Berkeley, Berkeley CA, 94720}
\affil[3]{School of Optometry and Vision Science, University of California, Berkeley, Berkeley CA, 94720}
\affil[4]{Chan Zuckerberg Imaging Institute, Redwood City CA, 94605}

\affil[*]{rcao@berkeley.edu}

\begin{abstract}
Structured illumination microscopy (SIM) uses a set of images captured with different illumination patterns to computationally reconstruct resolution beyond the diffraction limit. Here, we propose an alternative approach using a single speckle illumination pattern and relying on inherent sample motion to encode the super-resolved information in multiple raw images. From a set of raw fluorescence images captured as the sample moves, we jointly estimate both the sample motion and the super-resolved image. We demonstrate the feasibility of the proposed method both in simulation and in experiment.
\end{abstract}

\setboolean{displaycopyright}{false} 

\begin{document}

\maketitle

\section{Introduction}
Structured Illumination Microscopy (SIM) recovers a super-resolved image from a set of raw images captured with different illumination patterns (either sinusoidal~\cite{gustafsson2000surpassing} or speckle~\cite{mudry2012structured, yeh2019computational}). The structured illumination shifts the sample's higher-frequency content to lower spatial frequencies that are within the diffraction-limited bandwidth of the optical system. By capturing a set of raw images with different illumination patterns and solving a computational inverse problem, SIM can reconstruct up to 2$\times$ super-resolution. The varying illumination patterns serve as an encoding mechanism to ensure that diverse spatial frequency information is captured in the raw images. For sinusoidal SIM, \textasciitilde9 raw images are required, with illumination patterns having different orientations and phase shifts. For speckle SIM, dozens of raw images are needed for a well-posed inverse problem.

Updating the illumination pattern dynamically for each exposure causes SIM systems to be complicated and expensive. Furthermore, acquisition times can be long, especially if the optomechanical components are not as fast as the sensor framerate~\cite{fiolka2012time, forster2014simple,dan2013dmd}. This is particularly limiting for imaging live samples that move during the capture of the raw images, causing the reconstruction to suffer from motion artifacts. Recently, however, we developed a computational framework to estimate sample motion and eliminate motion artifacts for multi-shot imaging methods by modeling the motion with implicit neural representation~\cite{cao2024neural}. We applied it to several computational imaging systems, including a commercial sinusoidal SIM setup.

\begin{figure*}[tbh]
\centering
\includegraphics[width=0.86\linewidth]{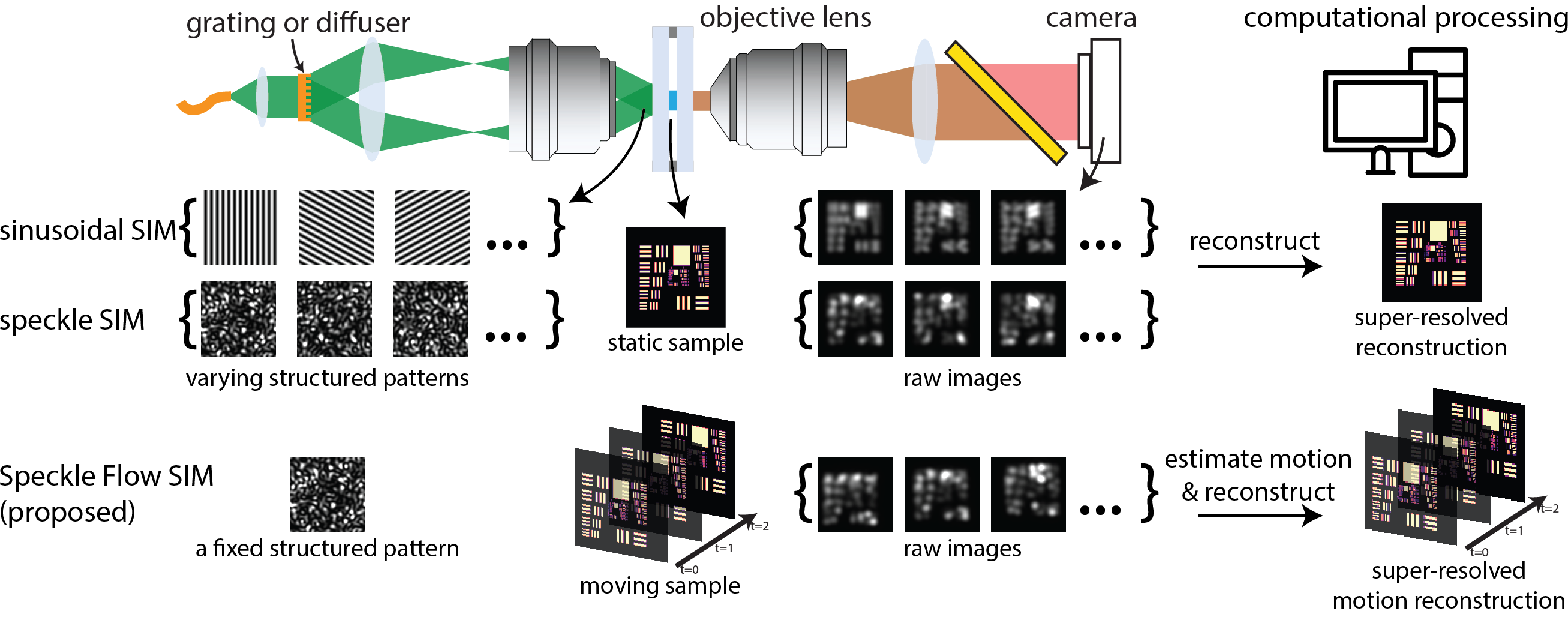}
\caption{Existing SIM methods (sinusoidal SIM and speckle SIM) vary the structured excitation patterns dynamically to encode super-resolution information into multiple raw images. We propose to implement SIM using a single fixed speckle illumination pattern and a moving sample, such that each part of the sample is illuminated by multiple different speckle patterns over time. Our algorithm then uses the captured images to solve for both the sample's motion and the super-resolved reconstruction.}
\label{fig:methods}
\end{figure*}

In this paper, we explore an alternative approach, called Speckle Flow SIM, that uses a single static illumination pattern during the acquisition of a sequence of raw images (Fig.~\ref{fig:methods}). Instead of varying the illumination over time, Speckle Flow SIM relies on sample motion to encode diverse spatial frequency information at different exposure times. Consider, for example, the case of a sample moving with translational motion. As the sample moves across the field-of-view, each part of the sample sees a different speckle illumination at each timepoint. By collecting a series of images as the sample moves, the super-resolution reconstruction problem becomes equivalent to that of having a laterally-shifting speckle illumination, which is straightforward to solve~\cite{yeh2019computational}. However, in our case, thanks to the neural network motion model, the motion may be unknown and arbitrary.

In order to accommodate unknown motion of the sample, we solve an inverse problem simultaneously for a sequence of super-resolved images, one for each timepoint in the captured raw data. This inverse problem is under-constrained because each raw image contains only a subset of the spatial frequencies required for the super-resolved frame. Nevertheless, it can be solved if we combine the information captured at all timepoints with motion estimation. To do so, we employ a neural space-time model (NSTM)~\cite{cao2022dynamic,cao2024neural} for flexibly modeling the spatiotemporal relationship of the moving sample. The NSTM assumes that the super-resolved image at one timepoint can be obtained by applying a motion transformation to the super-resolved image at a different timepoint. In essence, this means that the super-resolved video can be represented by a single super-resolved reference image as well as motion maps corresponding to each timepoint. The NSTM then simultaneously estimates the sample motion and the super-resolved image. While our Speckle Flow SIM idea has been previously demonstrated for coherent imaging~\cite{cao2022dynamic}, here we focus on fluorescence microscopy. 

We choose to use a speckle illumination pattern because it is non-directional and therefore suitable for achieving isotropic resolution gain. Sinusoidal illumination, in contrast, can only modulate high-frequency information along one orientation. Speckle patterns are also simple to generate by propagating a coherent beam through a random phase element, such as a diffuser. Unlike previous speckle SIM approaches that either rely on statistical properties~\cite{mudry2012structured} or require estimating the speckle pattern during reconstruction~\cite{min2013fluorescent,yeh2019computational}, our method's use of a static speckle pattern allows for pre-calibrating the pattern before the experiment. Therefore, our reconstruction algorithm assumes known illumination, requiring significantly fewer raw images (\textasciitilde10) than previous speckle SIM demonstrations (\textasciitilde50).

\section{Methods}
\subsection{SIM forward model}
SIM methods use patterned illumination, $I$, to excite a fluorescent sample, $S$. The emitted light is convolved with the diffraction-limited system's point-spread function (PSF) before being captured by the camera. The raw data, $D$, is expressed as
\begin{equation}
\label{eq:sim_forward}
\begin{split}
D \left( \bm{r} \right) &=  \left( I\left( \bm{r} \right) \cdot  S\left( \bm{r} \right) \right) \ast \text{PSF}\left( \bm{r} \right) 
= \mathcal{F}^{-1} \left[ \left(  \Tilde{I} \left( \bm{u} \right) \ast \Tilde{S} \left( \bm{u} \right) \right) \cdot \text{OTF}\left( \bm{u} \right) \right],
\end{split}
\end{equation}
\noindent where $\ast$ denotes convolution, $\cdot$ denotes Hadamard product (element-wise matrix product), $\mathcal{F}$, $\Tilde{\cdot}$ denote 2D Fourier transform, and $\bm{r}$ and $\bm{u}$ denote 2D coordinates in real and frequency space, respectively. The optical transfer function (OTF) is the Fourier transform of the PSF; here, it acts as a low-pass filter that describes the bandwidth allowed by the diffraction limit. 

If the illumination pattern contains any non-zero frequency content, then object frequencies beyond the diffraction-limited bandwidth will be encoded into $D$ due to the convolution between $\Tilde{I}$ and $\Tilde{S}$. Both sinusoidal and speckle patterns meet this condition; for example, sinusoidal illumination with a spatial frequency near the edge of the system passband results in capture of sample spatial frequencies up to 2$\times$ the diffraction limit. Using a sinusoidal pattern with a single orientation can only modulate frequencies along that orientation, so multiple orientations (and phase shifts) are needed to fully encode the super-resolution content. Speckle SIM achieves isotropic modulation since a speckle pattern provides more uniform coverage across all orientations. In both cases, the super-resolution information can be recovered computationally from the raw images(see Fig.~\ref{fig:methods}). 

\subsection{Speckle Flow SIM: Sample motion enabled SIM}
Speckle SIM encodes broad spatial-frequency information in each raw image, but still requires many raw images in order to disentangle spatial frequencies from the convolution in Eq.~\ref{eq:sim_forward}. Typically, this is achieved by varying speckle patterns over time. In Speckle Flow SIM, we need only a single static speckle pattern, and collect a sequence of raw images as the sample moves; thus, additional super-resolution information is encoded into the raw data at different timepoints. For example, in the simple case of a sample undergoing known translational motion with a relative displacement $\delta r$, the raw image can be expressed as:
\begin{equation}
\begin{split}
D \left( \bm{r} \right) &=  \left( I \left( \bm{r} \right) \cdot S\left( \bm{r} + \delta \bm{r} \right) \right) \ast \text{PSF}\left( \bm{r} \right) \\
&= \mathcal{F}^{-1} \left[ \left(  \Tilde{I} \left( \bm{u} \right) \ast \left( \Tilde{S} \left( \bm{u} \right) e^{-2\pi i \delta \bm{r} \cdot \bm{u}} \right) \right) \cdot \text{OTF}\left( \bm{u} \right) \right].
\end{split}
\end{equation}
The motion term, $e^{-2\pi i \delta \bm{r} \cdot \bm{u}}$, introduces additional frequency components as the displacement $\delta r$ changes, enriching the encoded super-resolution information. As more raw images are acquired at varying sample displacements, the super-resolved reconstruction becomes well-conditioned. The encoding process for non-linear deformable motion is more complicated, but can be approximated by multiple locally piecewise linear segments. Besides, the single static speckle pattern allows an easy pre-calibration, so that the pattern does not have to be solved during the reconstruction.

\subsection{Joint reconstruction of motion and sample}
Our neural space-time model (NSTM)~\cite{cao2024neural} has been shown capable of estimating unknown arbitrary deformable motion, as long as it is relatively smooth. NSTM models a dynamic scene using two multi-layer perceptrons (MLPs): a motion network, $f\left( \cdot, \theta_\text{m} \right)$, and a scene network, $f\left( \cdot, \theta_\text{s} \right)$. $\theta_\text{m}$, $\theta_\text{s}$ are the network weights. At each timepoint $t$, the moving sample is represented by $S\left( \bm{r}, t \right) = \text{NSTM} \left( \bm{r}, t ; \theta_\text{m}, \theta_\text{s} \right)$. With a set of raw images, $D_{t_1}, ..., D_{t_\text{n}}$ taken at timepoints $t_1, ..., t_\text{n}$, we solve for
\begin{equation}\label{eq:objective}
\begin{split}
\min_{\theta_{\text{m}}, \theta_{\text{s}}} \sum_t \| &D_t \left( \bm{r}, t \right) - \left( I \left( \bm{r} \right) \cdot \text{NSTM} \left( \bm{r}, t ; \theta_\text{m}, \theta_\text{s} \right) \right) \ast \text{PSF}\left( \bm{r} \right) \|^2. 
\end{split}
\end{equation}
We use gradient descent to jointly optimize the weights of the scene and motion networks.

\begin{figure}[tbh]
\centering
\includegraphics[width=\linewidth]{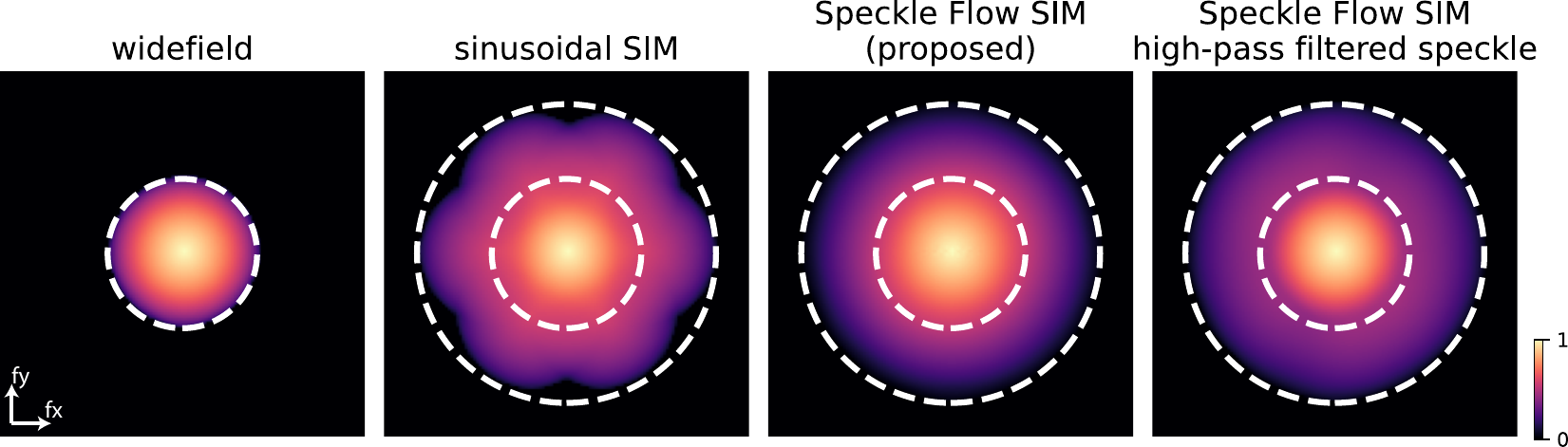}
\caption{Comparison of the effective transfer functions for widefield, sinusoidal SIM and Speckle Flow SIM (with and without high-pass filtering). The inner dashed white circle indicates the system's diffraction-limited bandwidth, and the outer circle indicates the theoretical 2$\times$ super-resolution bandwidth for SIM. Gamma corrections of 0.3 are applied to all for visibility.}
\label{fig:simu_encoding}
\end{figure}

\section{Simulation results}


To demonstrate the super-resolving capability of Speckle Flow SIM, we show two simulation studies: first, assessing the ability to encode super-resolution information during the forward process, and second, evaluating the performance of the inverse reconstruction process in decoding images. 

To assess encoding capabilities, we compare the forward process of our method with sinusoidal SIM. Since these methods require more than one raw image, we define an effective transfer function for each method that measures how much signal can be obtained for each possible input frequency. Mathematically, the effective transfer function, $T_{\text{eff}}\left(fx, fy\right) = \sum_t \| D_t\left( \delta \left( fx, fy\right) \right) - D_t\left( \bm{0} \right) \|$, where $\delta$ is an impulse at a particular spatial frequency and $D_t\left( \bm{0} \right)$ is the background raw image. Figure~\ref{fig:simu_encoding} shows the effective transfer functions for traditional widefield microscopy, sinusoidal SIM, and Speckle Flow SIM. Sinusoidal SIM uses nine raw images: three orientations and three phase shifts of the sinusoid per orientation. Speckle Flow SIM also uses nine raw images here, with the sample shifting laterally between images in this example case. Figure~\ref{fig:simu_encoding} demonstrates that Speckle Flow SIM indeed encodes information beyond the diffraction limit, though its encoding power at the highest spatial frequencies is weaker than sinusoidal SIM because the power of our speckle pattern is more concentrated in the lower frequency region. This limitation may be mitigated by shaping the speckles or high-pass filtering the speckle pattern.


\begin{figure}[tbh]
\centering
\includegraphics[width=\linewidth]{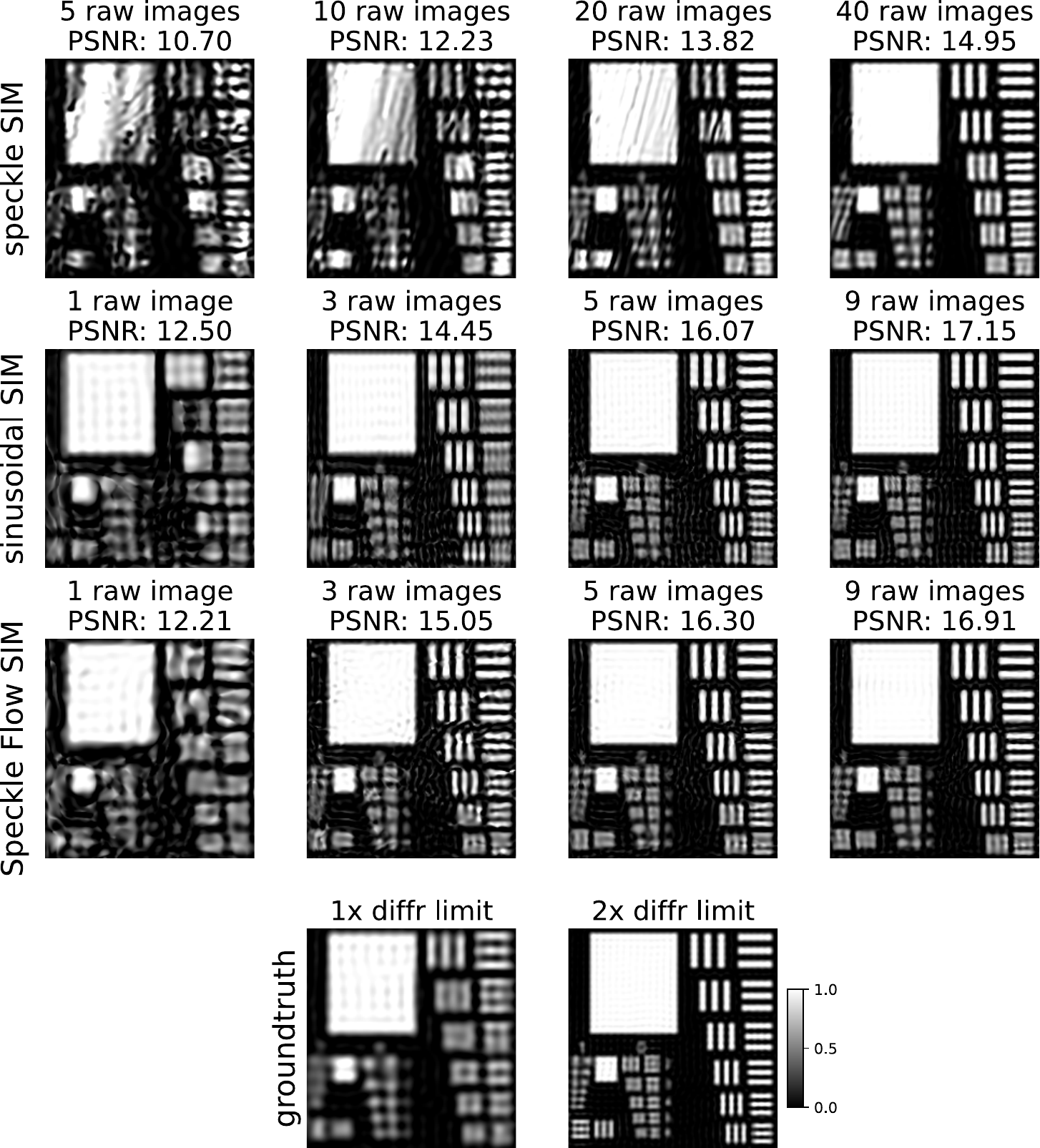}
\caption{Simulation comparison of speckle SIM, sinusoidal SIM and Speckle Flow SIM reconstructions with different numbers of raw images. The sample (USAF-1951 resolution target) is shifted laterally between raw images with known motion parameters for Speckle Flow SIM. Peak signal-to-noise ratio (PSNR) is reported for each reconstruction. Both sinusoidal SIM and Speckle Flow SIM use known illumination patterns and thus give much better reconstruction and data efficiency than speckle SIM. 
Speckle Flow SIM gives a better reconstruction for the cases with 3 raw images (with sinusoidal SIM using 3 phase-shifted images at a single orientation) and 5 raw images (sinusoidal SIM using 3 images at the first orientation and 2 images at the second orientation).}
\label{fig:simu_decoding}
\end{figure}

Next, we compared the decoding capabilities of Speckle Flow SIM with sinusoidal SIM and speckle SIM. All reconstructions used the same gradient descent algorithm and SIM forward model for super-resolution reconstruction from raw images, assuming the sample motion is known or has been estimated accurately. Figure \ref{fig:simu_decoding} shows the reconstruction with different numbers of raw images for each method. Due to the directional nature of sinusoidal patterns, the reconstruction quality of sinusoidal SIM suffers from anisotropic resolution when fewer raw images are captured. As shown in Fig.~\ref{fig:simu_decoding}, with three raw images, sinusoidal SIM fails to resolve any high-frequency vertical features as the sinusoidal patterns only modulate in the horizontal direction. In contrast, speckle patterns have isotropic frequency coverage, and thus are robust with a reduced number of raw images.

Similarly, we can compare the data efficiency between speckle SIM~\cite{yeh2019computational}, which jointly recovers the speckle patterns and the super-resolved image, and Speckle Flow SIM, which uses a known speckle pattern, in Fig.~\ref{fig:simu_decoding}. Speckle Flow SIM requires significantly fewer raw images than speckle SIM to obtain a high-quality reconstruction, because of the a priori knowledge of the speckle. Even with only 9 raw images, Speckle Flow SIM achieves better reconstruction than speckle SIM using 40 images.

\section{Experimental results}
We built a transmission mode setup to experimentally demonstrate the concept of Speckle Flow SIM. A 488nm laser (Coherent OBIS LS) serves as the light source. As in Fig.~\ref{fig:methods}, the beam passes through a beam expander (Thorlabs GBE05-A), and then the speckle pattern is generated by light interference after two layers of Scotch tape. This speckle pattern excites the sample and the emitted light is magnified by a 4-$f$ system consisting of a 10$\times$ 0.25NA objective (Nikon Plan) and a tube lens (250mm focal length). The excitation light is filtered by a dichroic mirror and an emission band-pass filter (Edmund Optics 67-079 and 86-984), before the sensor (PCO Edge 5.5) collects the fluorescent signal. To calibrate the system, the excitation pattern is pre-captured without the sample and the emission band-pass filter in place.

\begin{figure}[tbh]
\centering
\includegraphics[width=0.96\linewidth]{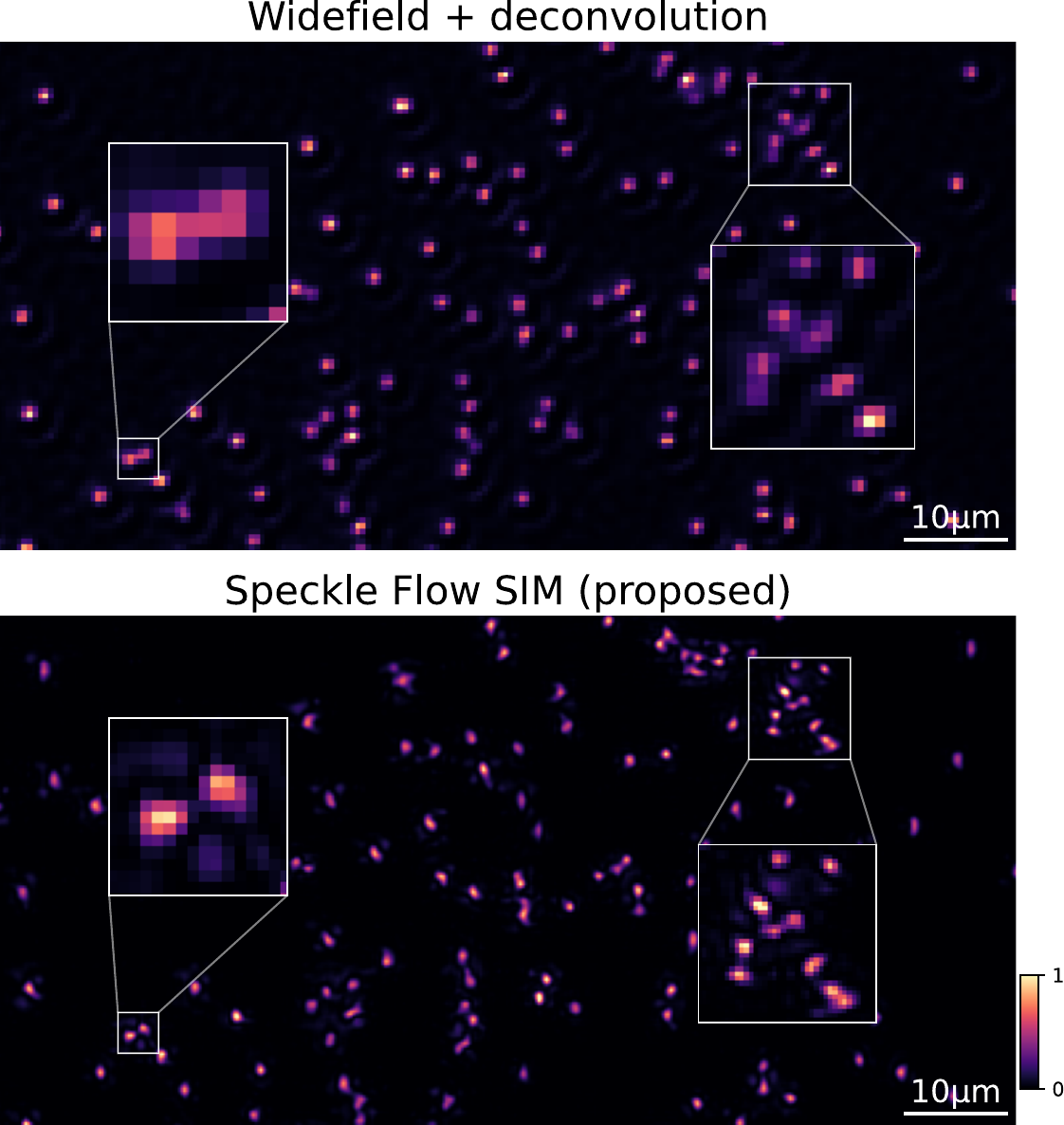}
\caption{Experimental result of Speckle Flow SIM on a fluorescence-labeled microbead sample. Top: Reconstruction with widefield excitation and image deconvolution using the theoretical system PSF. Bottom: One timepoint from the reconstruction by Speckle Flow SIM, demonstrating super-resolution and accurate motion estimation (full video in Visualization 1). During the acquisition, the sample moves continuously on a motorized lateral stage.}
\label{fig:exp_result}
\end{figure}

A fluorescence-tagged sample with 0.71µm-diameter microbeads (Thermo Scientific) is placed on a motorized lateral stage (Thorlabs PT1-Z8). During acquisition, a sequence of raw images is captured as the lateral stage continuously moves the sample. Although the sample motion is controlled, the stage motion is unknown to the reconstruction algorithm. NSTM is used to jointly solve for the motion and the scene by optimizing the objective defined in Eq.~\ref{eq:objective}. We run gradient descent optimization for 10,000 epochs with an initial learning rate of 0.0001 for both the motion and scene networks, and the learning rate has a decay of 0.8 every 1,000 epochs. 

Figure~\ref{fig:exp_result} presents the reconstruction result from the Speckle Flow SIM experimental system and compares it with the widefield image after PSF deconvolution. Speckle Flow SIM successfully resolves individual microbeads that the widefield image fails to capture, demonstrating that the contrast introduced by sample motion enables super-resolution. With the joint motion and scene reconstruction from NSTM, Speckle Flow SIM recovers the dynamic super-resolved images for a moving sample with unknown motion. The full dynamic super-resolution reconstruction is in Supplemental Video 1.

\section{Discussion}
Sample motion between captured images has traditionally been viewed as a problem for super-resolution microscopy, causing motion blurring artifacts. Here, we have utilized it as a novel contrast mechanism for SIM. Rather than relying on time-varying excitation patterns, we rely on sample motion to encode diverse information into the measurements with a fixed excitation pattern. This greatly simplifies the implementation of SIM systems and removes motion blurring artifacts. With a fixed pattern, optomechanical components or spatial light modulators are no longer needed to update the excitation patterns, consequently reducing the equipment cost as well as eliminating the time delay for the pattern update. Furthermore, a single fixed speckle pattern is easy to pre-calibrate, unlike traditional speckle SIM, which enhances the data efficiency of the reconstruction.

A limitation of our work is that it requires motion to achieve super-resolution. For static parts of the sample, the raw data will encode identical information over all exposures, resulting in an ill-posed inverse problem. This can be mitigated by introducing external global motion, like shifting the sample laterally.

The proposed motion contrast mechanism is enabled by recent advances in dynamic imaging reconstruction, allowing us to jointly estimate both the motion and the scene. This joint estimation relies on a temporal redundancy assumption, which means that the scene at any given timepoint is highly correlated with the scene at the previous timepoint. However, we want to note that this assumption does not always hold true as discussed in~\cite{cao2024neural}, potentially causing the failure of motion estimation. This limitation restricts the use of Speckle Flow SIM to samples exhibiting smooth motion, such as cells in microfluidic channels, rather than samples with abrupt, on/off motion like neuron firing. In future research, other types of motion assumptions and modelings can be explored to alleviate this motion constraint.

If motion is very fast, then individual raw images may become blurred. Such motion blurring is not accounted for by our current algorithm; it could lead to information loss during encoding and cause model mismatch in motion estimation. Therefore, it is essential to carefully select the exposure settings to minimize blur in the raw images, which is a common requirement in other microscopy methods. Since the NSTM can predict motion during each raw image's exposure time, future work could attempt to deblur the raw images computationally.

In summary, Speckle Flow SIM employs a static speckle illumination pattern to encode super-resolution information in fluorescence imaging. We demonstrated its feasibility through simulation analysis and proof-of-concept experiments. In our experiments, Speckle Flow SIM achieved super-resolution reconstruction on a continuously moving fluorescent-labeled sample. This motion contrast mechanism creates new opportunities for innovative optical designs for SIM systems.

\begin{backmatter}
\bmsection{Funding} Chan Zuckerberg Initiative (DAF2021-225666, DAF2021-225666 (DOI: 10.37921/192752jrgbnh)); Chan Zuckerberg Initiative (Biohub SF investigator (L.W.)); National Science Foundation (DMR 1548924); Siebel Scholars Foundation (Siebel Scholar (R.C.)).


\bmsection{Disclosures} The authors declare no conflicts of interest.

\bmsection{Data Availability Statement} 
Data and code in this paper are available in \href{https://github.com/rmcao/SpeckleFlowSIM-fluo}{https://github.com/rmcao/SpeckleFlowSIM-fluo} .

\end{backmatter}

\bibliography{main}

\begin{thebibliography}{1}
\newcommand{\enquote}[1]{``#1''}

\bibitem{gustafsson2000surpassing}
M.~G. Gustafsson, \enquote{Surpassing the lateral resolution limit by a factor of two using structured illumination microscopy,} {\protect\JournalTitle{Journal of microscopy}} \textbf{198}, 82--87 (2000).

\bibitem{mudry2012structured}
E.~Mudry, K.~Belkebir, J.~Girard, \emph{et~al.}, \enquote{Structured illumination microscopy using unknown speckle patterns,} {\protect\JournalTitle{Nature Photonics}} \textbf{6}, 312--315 (2012).

\bibitem{yeh2019computational}
L.-H. Yeh, S.~Chowdhury, and L.~Waller, \enquote{Computational structured illumination for high-content fluorescence and phase microscopy,} {\protect\JournalTitle{Biomedical Optics Express}} \textbf{10}, 1978--1998 (2019).

\bibitem{fiolka2012time}
R.~Fiolka, L.~Shao, E.~H. Rego, \emph{et~al.}, \enquote{Time-lapse two-color 3d imaging of live cells with doubled resolution using structured illumination,} {\protect\JournalTitle{Proceedings of the National Academy of Sciences}} \textbf{109}, 5311--5315 (2012).

\bibitem{forster2014simple}
R.~F{\"o}rster, H.-W. Lu-Walther, A.~Jost, \emph{et~al.}, \enquote{Simple structured illumination microscope setup with high acquisition speed by using a spatial light modulator,} {\protect\JournalTitle{Optics Express}} \textbf{22}, 20663--20677 (2014).

\bibitem{dan2013dmd}
D.~Dan, M.~Lei, B.~Yao, \emph{et~al.}, \enquote{Dmd-based led-illumination super-resolution and optical sectioning microscopy,} {\protect\JournalTitle{Scientific Reports}} \textbf{3}, 1116 (2013).

\bibitem{cao2024neural}
R.~Cao, N.~S. Divekar, J.~K. Nu{\~n}ez, \emph{et~al.}, \enquote{Neural space--time model for dynamic multi-shot imaging,} {\protect\JournalTitle{Nature Methods}} \textbf{21}, 2336--2341 (2024).

\bibitem{cao2022dynamic}
R.~Cao, F.~L. Liu, L.-H. Yeh, and L.~Waller, \enquote{Dynamic structured illumination microscopy with a neural space-time model,} {\protect\JournalTitle{2022 IEEE International Conference on Computational Photography (ICCP)}} pp. 1--12 (2022).

\bibitem{min2013fluorescent}
J.~Min, J.~Jang, D.~Keum, \emph{et~al.}, \enquote{Fluorescent microscopy beyond diffraction limits using speckle illumination and joint support recovery,} {\protect\JournalTitle{Scientific Reports}} \textbf{3}, 2075 (2013).

\end{thebibliography}


\end{document}